\documentclass[runningheads,a4paper]{llncs}
\usepackage{listings}
\usepackage{amssymb}
\usepackage{graphicx}
\usepackage{pgfplots}
\usepackage{longtable}
\usepackage{multirow}
\usepackage[ruled,vlined]{algorithm2e}
\DeclareGraphicsExtensions{.eps}
\usepackage{url}    
\newcommand{\keywords}[1]{\par\addvspace\baselineskip
\noindent\keywordname\enspace\ignorespaces#1}

\graphicspath{{pictures/}}
\DeclareGraphicsExtensions{.png,.jpg}

\lstdefinestyle{customasm}{
  belowcaptionskip=1\baselineskip,
  frame=L,
  xleftmargin=\parindent,
  language=[x86masm]Assembler,
  basicstyle=\tiny\ttfamily,
  commentstyle=\itshape\color{purple!40!black},
}
\begin{document}

\mainmatter  

\title{Improving fuzzing using software complexity metrics}

\titlerunning{Improving fuzzing using software complexity metrics}

%
%
\author{Maksim O. Shudrak, Vyacheslav V. Zolotarev}
\authorrunning{Improving fuzzing using software complexity metrics}

\institute{IT Security Department, Siberian State Aerospace University,\\
	Krasnoyarsky Rabochy. Av. 31, 660014 Krasnoyarsk, Russia\\
	mxmssh@gmail.com, amida@land.ru
}

%
%

\toctitle{Improving fuzzing using software complexity metrics}
\maketitle

\begin{abstract}
Vulnerable software represents a tremendous threat to modern information systems. Vulnerabilities in widespread applications may be used to spread malware, steal money and conduct target attacks. To address this problem, developers and researchers use different approaches of dynamic and static software analysis; one of these approaches is called fuzzing. Fuzzing is performed by generating and sending potentially malformed data to an application under test. Since first appearance in 1988, fuzzing has evolved a lot, but issues which addressed to effectiveness evaluation have not fully investigated until now.\newline
In our research, we propose a novel approach of fuzzing effectiveness evaluation and improving, taking into account semantics of executed code along with a quantitative assessment. For this purpose, we use specific metrics of source code complexity assessment specially adapted to perform analysis of machine code. We conducted effectiveness evaluation of these metrics on 104 wide-spread applications with known vulnerabilities. As a result of these experiments, we were able to identify the best metrics that is more suitable to find bugs. In addition we proposed a set of open-source tools for improving fuzzing effectiveness. The experimental results of effectiveness assessment have shown viability of our approach and allowed to reduce time costs for fuzzing campaign by an average of 26-28\% for 5 well-known fuzzing systems.

\end{abstract}

\section{Introduction}

Nowadays each software product should meet a number of conditions and requirements to be useful and successful on the market. Despite this fact, software engineers and developers keep making mistakes (bugs) during software development. In turn, these bugs can create favorable conditions for emergence of serious vulnerabilities. This is particularly relevant for network applications because vulnerabilities in this type of software create great opportunities for an attacker, such as remote code execution or DoS attack. However, practice has shown that vulnerabilities in local applications may also present a serious threat to information systems if they allow to execute arbitrary code in the context of а vulnerable application. This severely endangers commercial success of the product and can considerably decrease the security rate of infrastructure as well. Critical vulnerabilities in widespread products deserve special attention because they are often a target for mass malware attacks and persistent threats. Suffice it to say that in 2014, US National Vulnerability Database registered 26 new vulnerabilities per day on average \cite{L1}.

There are two fundamentally different approaches for bugs detection in binary executables: static and dynamic analysis. Static analysis is aimed at finding bugs in applications without execution, while dynamic analysis performs bugs detection at runtime.

In our research, we consider only binary code of the program. Binary code (machine code, executable code) is a code (a set of instructions) executed directly by a CPU. The reason of this is due to the presence of proprietary software that is distributed in binary form only. The second problem related to transformations performed by compilers and optimizer tools that may significantly change actual behavior of the program in the binary form. This problem is called «What You See Is Not What You eXecute» \cite{L2}.

In the paper, we will use technique of dynamic analysis called fuzzing. Fuzzing is performed by generating and sending potentially malformed data to an application. The first appearance of fuzzing in software testing dates back to 1988 by Professor Barton Miller \cite{L7}; since then the fuzzing has evolved a lot and it is now used for vulnerabilities detection and bugs finding in a large number of different applications. There are a lot of instruments for fuzzing, such as Sulley \cite{L3}, Peach \cite{L4}, SAGE \cite{L5} and many others. However, issues which addressed to effectiveness evaluation have not fully investigated until now.

Today researchers often use several basic criteria for effectiveness evaluation: the number of errors found, the number of executed instructions, basic blocks or syscalls as well as cyclomatic complexity or attack surface exposure \cite{L6}--\cite{L9}.

During the last several decades, the theory of software reliability has proposed a wide range of different metrics to assess source code complexity and the probability of errors. The general idea of this assessment is that more complex code has more bugs. In this paper, our hypothesis is that source code complexity assessment metrics could be adapted to use them for binary code analysis. Thus it would allow to perform analysis based on semantics of executed instructions as well as their interaction with input data.

We will provide an overview of technique, architecture, implementation, and effectiveness evaluation of our approach. We will carry out separate tests to compare effectiveness of 25 complexity metrics on 104 wide-spread applications with known vulnerabilities. Moreover, we will perform assessment of our approach to reduce time costs of fuzzing campaigns for 5 different well-known fuzzers.

The purpose of this research was to increase effectiveness of the fuzzing technique in general, regardless of the specific solutions. Thus, we did not develop our own fuzzer, but focused on flexibility of our tools by making them easy to use with any fuzzers. Thus we did not try to improve test cases generation or mutation to find more bugs but we try to make fuzzing campaign more efficient in terms of time costs required to detect bugs in software.

The contributions of this paper are the following:
\begin{enumerate} 
\item	We adapted a set of source code complexity metrics to perform fuzzing effectiveness evaluation by estimating complexity of executable code.
\item	We conducted the comparative experimental evaluation of proposed metrics and identified the most appropriate ones to detect bugs in executable code.
\item	We implemented a set of tools for executable code complexity evaluation and executable trace analysis. In addition, we also made our tools and experimental results accessible for everyone in support of open science \cite{L28}.
\end{enumerate}
The paper is structured as follows. In Section 2 we illustrate short overview of fuzzing and problems of its effectiveness evaluation. Section 3 covers details of metrics adaptation. Then, Section 4 provides an in-depth description of system implementation. Detailed results of metrics effectiveness evaluation and their comparison are presented in Section 5. Section 6 used to present experimental results of system integration with well-known fuzzers. Further, we outline related works in Section 7 and describe the direction of our future research in Section 8. Finally, we use Section 9 to present conclusions.

\section{Problem Statement}

In Section 1, we mentioned that fuzzing is performed by generating and sending potentially malformed data to an application. Nowadays, fuzzing is used for testing different types of input interfaces such as: network protocols \cite{L10}, file formats \cite{L11}, in-memory fuzzing \cite{L12}, drivers and many others software and hardware products that process input data. Moreover, fuzzing is not limited to pseudorandom data generation or mutation, but includes a mature formal data description protocol and low-level analysis of binary code for generating data and monitoring  results. However, the question still remains: \emph{“How can we evaluate fuzzing effectiveness?”} Of course, we can assess it by the number of bugs detected in an application. But this is not a flexible approach, since it does not provide any information on how well the testing data was generated or mutated in case when the analysis showed no errors at all. On the other hand, for this purpose, we can use code coverage, assuming that the higher is code coverage, the more effective the testing. Code coverage is a measure used to describe the degree to which the code of a program is tested by a particular test suite. In most cases, researchers assess code coverage by calculating the total number of instructions, basic blocks or routines that have been executed in the application under test. However, they do not take into account the complexity of tested code. For example, different code paths may have equal values of code coverage but their complexity may be different. Let us consider the example in Figure 1.

\begin{figure}
\centering
\lstset{language={[x86masm]Assembler}}
\noindent\begin{minipage}{.24\textwidth}
\begin{lstlisting}[title = Listing A]{Name}
push eax
push 0Ah
lea eax, [ebp+Source]
push eax
call fgets
add esp, 0Ch
lea eax, [ebp+Source]
push eax
lea ecx, [ebp+Format]
push ecx
call strcpy
add esp, 8
cmp [ebp+var_34], 0
jnz short loc_4135B6
\end{lstlisting}
\end{minipage}
\begin{minipage}{.24\textwidth}
\begin{lstlisting}[title = Listing B]{Name}
push eax
push offset Format
call scanf
add esp, 8
mov eax, [ebp+b]
imul eax, 6
add eax, 3
mov ecx, [ebp+b]
imul ecx, 6
add ecx, 3
imul eax, ecx
add eax, [ebp+a]
mov [ebp+a], eax
jnz short loc_4135AD
\end{lstlisting}
\end{minipage}
\caption{Two different code blocks with equal code coverage measure}
\end{figure}
The code in Listing A handles user data and may contain buffer overflow, whereas the code in Listing B reads an integer and performs some calculations by using this value. Code coverage for these examples is the same, but the code in Listing A is more interesting for analysis.

Basili V. \cite{L13}, Khoshgoftaar M. \cite{L14}, Olague H. \cite{L15} and other researchers have shown that in general, increasing of code complexity leads to increase in the probability of an error. This contention is supported by experimental results \cite{L6}--\cite{L9}. 

In this paper, we propose to adapt source code complexity assessment metrics so as to take into account semantics of binary code. We propose the following hypothesis: \emph{"There is a more effective complexity metric for fuzzing effectiveness assessment than the number of executed instructions, basic blocks and routines, as well as than cyclomatic complexity"}.
Thus, we need to adapt complexity metrics for binary code and then perform analysis of their effectiveness in comparison with traditional metrics. 

In our research, we consider the following types of errors: buffer and heap overflows, format string errors, read and write to invalid or incorrect memory address, null pointer dereferences, use after free, as well as use of uninitialized memory.

\section{Metrics adaptation}

In the article, we adapted 25 metrics of source code complexity assessment. Without getting into description of each metric, let us describe symbols and references to the authors of each measure.

\begin{itemize} 
\item Lines of code count (\emph{LOC}), basic blocks count (\emph{BBLs}), procedure calls count (\emph{CALLS)};
\item Jilb metric (\emph{Jilb}) \cite{L16}, ABC metric (\emph{ABC}), Cyclomatic complexity (\emph{CC}) \cite{L17}, Modified cyclomatic complexity (\emph{CC\_mod})\cite{L16}, density of CFG (\emph{R}) \cite{L18}, Pivovarsky metric (\emph{Pi}) \cite{L16}, Halstead metrics for code volume (\emph{H.V}), length and calculated length (\emph{H.N, $H.N^*$}), difficulty (\emph{H.D}), effort (\emph{H.E}), the number of delivered bugs (\emph{H.B}) \cite{L19};
\item Harrison and Magel metric (\emph{Harr}) \cite{L20}, boundary values metric (\emph{Bound}), span metric (\emph{Span}), Henry and Cafura metric (\emph{H\&C}) \cite{L21}, Card and Glass metric (\emph{C\&G}) \cite{L22}, Oviedo metric (\emph{Oviedo}) \cite{L23}, Chapin metric (\emph{Chapin}) \cite{L24};
\item Cocol metric (\emph{Cocol}) \cite{L16}.
\end{itemize}

The detailed description of each adapted metric is also given in the Appendix 1. Metrics that take into account high level information such as source code comments, name of variables or some object oriented information were excluded from the scope of this analysis.

It should be noted that for most of the metrics we need to perform conversion of routines code into control flow graph (CFG). CFG has only one entry and one exit. A path in the CFG can be represented as an ordered sequence of node numbers. In terms of binary code analysis, graph nodes are represented as a basic block of instructions and edges describe control flow transfer between basic blocks. Basic block (linear block) is a set of machine instructions without conditional or unconditional jumps excluding function calls. Algorithm 1 allows to perform such conversion.

\begin{algorithm}
 \SetAlgorithmName{Algorithm}{}
 \DontPrintSemicolon
 \KwData{Address of the first instruction, an empty set of links}
 \KwResult{A set of nodes, A set of edges }
 \BlankLine
 \While{not end of routine}{
  Read instruction\;
  \If{First instruction in the node}{
   Save instruction address as the first address of the node\;
   }
   Get links of the instriction\;
  \If{Number of links $>$ 0}{
  	Save instruction address as the last address of the node\;
  	Save edges in a set of edges\;
  }   
  Move the pointer to the next instruction\;
 }
 \caption{Routine to CFG translation}
\end{algorithm}

The algorithm passes through all basic blocks in the routine. A link is conditional or unconditional jump to some address within routine code. Note that the link is not considered for call instructions. Each instruction at some address may have from 0 up to n outgoing links. Unconditional jump always has two links, the first one refers to the address of unconditional jump, and the second one is the link to the address following immediately after jump instruction. Thus each node is associated with the following information: address of the head, address of the end, edge address 1 (optional) and edge address 2 (optional).

Note that bugs may arise from the use of unsafe library functions, such as $strcpy$, $strcat$, $lstrcat$, $memcpy$ and etc. These functions are banned or not recommended to use, since they may cause overflows in the memory. Efficient fuzzing campaign should take into account this fact and firstly cover the routines that call these functions. In the article, we propose to use the following experimental measure based on Halstead B metric (rationale for the choice of this metric is proposed in Section 5):
\begin{equation}\label{eq:expmetric}
Exp = H.B \times \sum\limits_{i=1}^{n} (v_i + 1)
\end{equation}
n - a total number of banned or not recommended functions used in the routine. $v_i$ is calculated as the total number of calls of banned or not recommended functions in the routine, multiplied by the coefficient of the potential danger associated with this syscall. This coefficient calculated by using the banned functions list proposed by Microsoft within their secure development lifecycle concept \cite{L25}. In our research, a function can take only two values: 0.5 for dangerous and 1 for banned syscalls. It should be noted that multiplication is used to prioritize routines that calls unsafe functions.

\section{System overview}
\subsection{Fuzzing strategy}

Let's describe all basic blocks in a program as an ordered set of nodes: $CFG$ = \{$node_0$, $node_1$,...,$node_n$\}, where $node$ is a basic block and $n$ - total number of basic blocks. Let's define an array of test data as $TD$ = [$td_0$, $td_1$,...,$td_\nu$], $\nu$ - an array size and $td$ - one instance of test data (file, network packet, etc.) to make one fuzzing iteration. Then code coverage for one test iteration may be written as:

\begin{equation}\label{eq:cover}
Cover = [cov_0, ... , cov_\nu]
\end{equation}

Then, let's assign weight for each test case and sort them in descending order of weight. Weights for test cases is assigned using complexity of trace which is calculated using metrics described above. Further we will send test cases according with their position in the sorted array.

In the case of adding new test data in TD without associated coverages, new instances take the highest priority with respect to existing elements, and passed to the program in random order before existing test cases.

\subsection{Trace analysis}

As it was noted in the second section, we need to save addresses of instructions, basic blocks or routines to assess complexity of code that has been executed during analysis. In this research, we used technique called dynamic binary instrumentation to perform code coverage analysis. Dynamic Binary Instrumentation (DBI) is a technique of analyzing the behavior of a binary application at runtime through the injection of instrumentation code. The main advantage of DBI is the ability to perform binary code instrumentation without switching the processor context, which significantly improves performance. In our research we use DBI framework called Pin \cite{L26}. Pin provides API to create the dynamic binary analysis tools called PinTools. Pin performs dynamic translation of each instruction and adds instrumentation code, if it required. Note that dynamic translator performs code translation without intermediate stages in the same architectures (IA32 to IA32, ARM to ARM and etc).

\subsection{Metrics evaluation module}

Let us describe basic scheme of the tool for binary code complexity assessment in Figure 2.

\begin{figure}
\centering
\center{\includegraphics[scale=0.30]{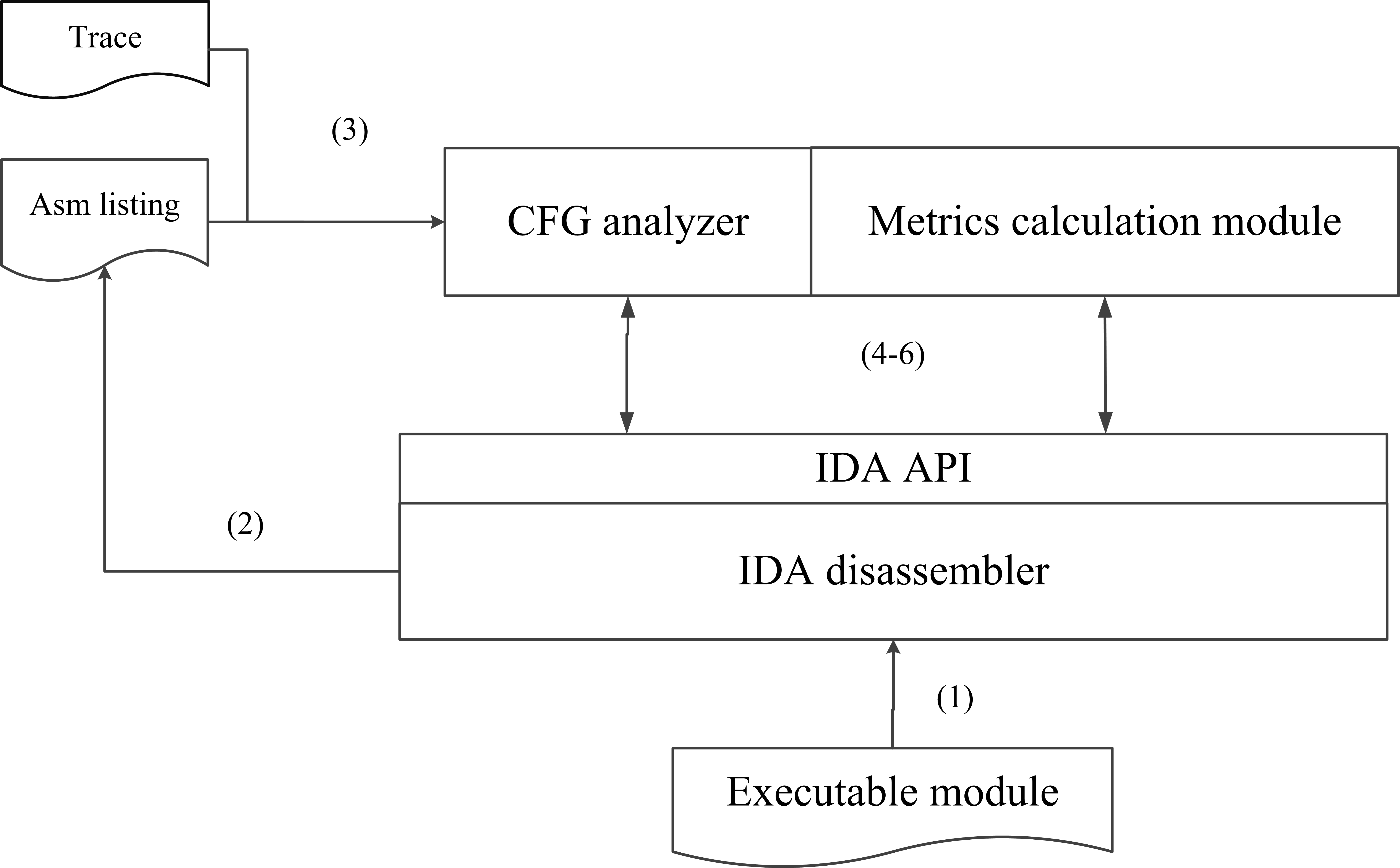}}
\caption{Scheme of the tool for binary code complexity assessment}
\end{figure}

 At the first stage, we use IDA disassembler to perform preliminary analysis and disassembling of executable module. Then assembler listing and trace is passed to the module of CFG analysis that sequentially iterates through each executed basic block in the program. The routine parser performs analysis of interconnections between basic blocks on the basis of which the tool builds graph of a routine. This graph is used in the module of metrics calculation that performs analysis and evaluation of each complexity measure for each required metric. Where necessary, this module also uses the binary code translation to get information required for some metrics. For example, the total number of assignments could be in turn obtained by using high level listing obtained by the translator, where operations like $eax$ = $eax + 1$ may be considered as an assignment.
 
\section{Metrics effectiveness evaluation}

In section 2, it was mentioned that we need to perform effectiveness comparison between adapted and traditional metrics. To meet this challenge, we decided to use open database with vulnerable applications called exploit-db supported by Offensive Security \cite{L27}. In our experiment, we randomly selected 104 different vulnerable applications. This is minimum sample size which is required to evaluate the effectiveness of all the metrics in the 95\% confidence interval within an error no more than 3\%. As a result we randomly selected the following types of applications: video and audio players; FTP, HTTP, SMTP, IMAP and media servers; network tools; scientific applications; computer games; auxiliary tools (downloaders, torrent-clients, development tools and etc.); libraries (converters, data parsers and etc.); readers (PDF, DJVU, JPEG and etc.); archivers and etc. For details please visit \cite{L28}.

Then exploit has been found for each program which allowed to locate vulnerable routine in the application. Each application was in turn analyzed by the tool of code complexity assessment. Complexity of each metric has been obtained for each routine in each vulnerable application. Then obtained measures were ranked in descending order. Lastly, we selected ranks of all vulnerable routines in each application (The results for each application may be found at \cite{L28}). Obviously that obtained results do not allow to assess and compare effectiveness of metrics, since they do not take into account total number of routines in the application.\emph{An effective metric is a metric that takes a maximum complexity value for vulnerable routines.} Thus the following formula was used to solve this problem:

\begin{equation}\label{eq:PR}
PR = (1-\frac{frang}{TF}),
\end{equation}

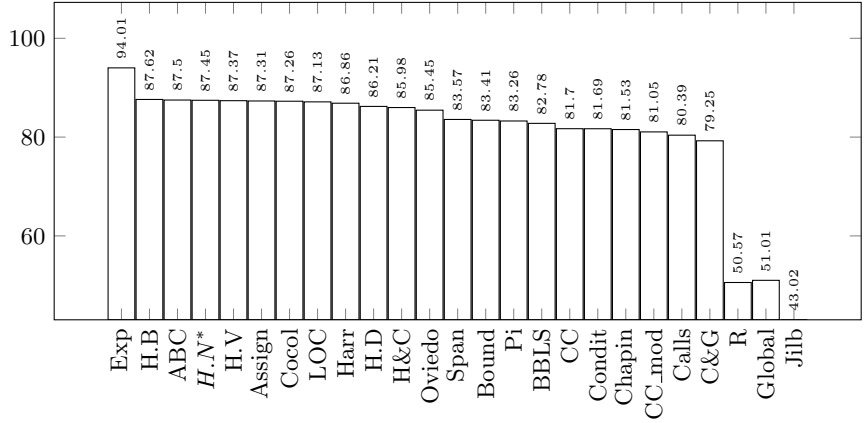
\begin{figure}[tb]
    \centering
	\begin{tikzpicture}
	\begin{axis}[
	symbolic x coords={Exp,H.B, ABC, $H.N^*$, H.V,Assign,Cocol,LOC,Harr,H.D, H\&C,Oviedo,Span,Bound,Pi,BBLS,CC,Condit,Chapin,CC\_mod, Calls,C\&G,R,Global,Jilb},
	xtick=data,
    enlarge y limits={upper,value=0.26},
    x tick label style={rotate=90},
    nodes near coords,
    every node near coord/.append style={rotate=90, font=\tiny},
    nodes near coords align={horizontal},
    width=\linewidth,
    height = 0.3\textheight
    ]
	\addplot[ybar] coordinates {
		(Exp, 94.01) (H.B, 87.62) (ABC, 87.50) ($H.N^*$, 87.45) (H.V, 87.37)
        (Assign, 87.31) (Cocol, 87.26) (LOC, 87.13) (Harr, 86.86) (H.D, 86.21)
        (H\&C, 85.98) (Oviedo, 85.45) (Span, 83.57) (Bound, 83.41) (Pi, 83.26)
		(BBLS, 82.78) (CC, 81.70) (Condit, 81.69) (Chapin, 81.53)
		(CC\_mod, 81.05) (Calls, 80.39) (C\&G, 79.25) (Global, 51.01)
        (R, 50.57) (Jilb, 43.02)
	};
	\end{axis}
	\end{tikzpicture}
	\caption{Average effectiveness of each metric. Experimental metric demonstrates maximum effectiveness. Y - percent interval.}
\end{figure}

$frang$ - a routine rank and $TF$ - a total number of routines. This expression enables to answer the following question: \emph{“How many routines in a program have metric values less than for a vulnerable routine?”} This value in percent may be obtained for each metric in each application. Now, it's possible to calculate average measures for each metric (Figure 3).

According to Figure 3, $Jilb$, $Global$ and $R$ metrics showed the lowest average values. Let's exclude these metrics from further analysis. Also, it makes sense to exclude $H.D$, $H.V$ and $H.N^*$ metrics, since they're used to calculate $H.B$ and showed comparable results. 

Let us compare metrics using coefficient of variation (Figure 4). Coefficient of variation is used to show the extent of variability in relation to the mean of the value.

\begin{figure}
    \centering
	\begin{tikzpicture}
	\begin{axis}[
	symbolic x coords={Exp,H.B,ABC,Assign,Cocol,LOC,Harr,H\&C,Oviedo,Span,Bound,Pi,BBLS,CC,Condit,Chapin,CC\_mod, Calls,C\&G,R,Global,Jilb},
	xtick=data,
	ymin=5,
    enlarge y limits={upper,value=0.26},
    x tick label style={rotate=90},
    nodes near coords,
    every node near coord/.append style={rotate=90, font=\tiny},
    nodes near coords align={horizontal},
    width=\linewidth,
    height = 0.3\textheight
    ]
	\addplot[ybar] coordinates {
		(Exp, 9.39) (H.B, 16.31) (ABC, 16.58) (Assign, 16.82) 
		(Cocol, 17.52) (LOC, 17.85) (Harr, 17.79) (H\&C, 17.23)
		(Oviedo, 21.32) (Span, 25.83) (Bound, 26.81) (Pi, 27.09)
		(BBLS, 27.31) (CC, 30.33) (Condit, 29.54) (Chapin, 30.14)
		(CC\_mod, 31.79) (Calls, 34.34) (C\&G, 35.15)
	};
	\end{axis}
	\end{tikzpicture}
	\caption{Coefficients of variation for metrics (less is better). Y axis - coefficient of variation. Cyclomatic complexities, Chapin and Card\&Glass demonstrate high level of variation.}
\end{figure}
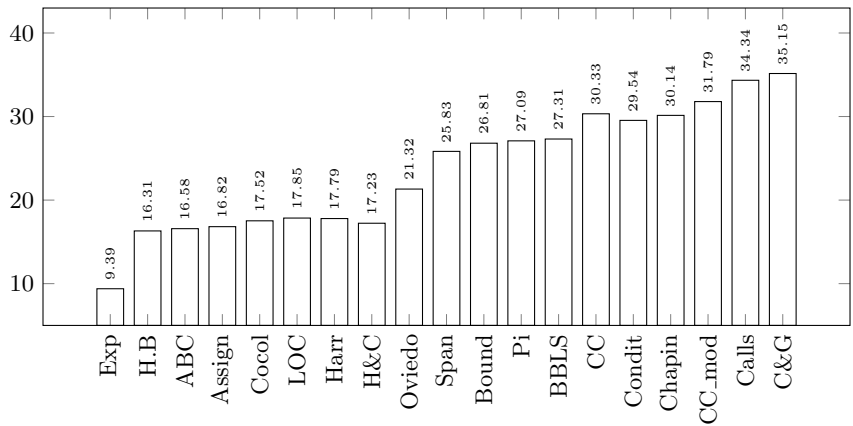

The obtained statistical results have shown that experimental metric exceeds metrics based on cyclomatic complexity (by 12,31\%) the number of basic blocks (by 11,23\%), calls (by 13,62\%), LOC (6.88\%) and at the same time has the lowest coefficient of variation 9.4\%. Note that the statistical error for the experimental metric is ±2,54\% at 95\% confidence interval. Thus, all of these data prove that hypothesis proposed in the section 2 is correct.

In section 3 it was noted that the basis of experimental metric is Halstead B measure. We use this measure because Halstead B demonstrated the best effectiveness compared to other known metrics.

\section{Experiments}
\subsection{Code coverage analysis}

According to section 5, the system is based on 2 modules: module of metrics calculation and module of trace analysis. Let's describe the general scheme of the system integration with fuzzer in Figure 5. 

\begin{figure}
\centering
\center{\includegraphics[scale=0.3]{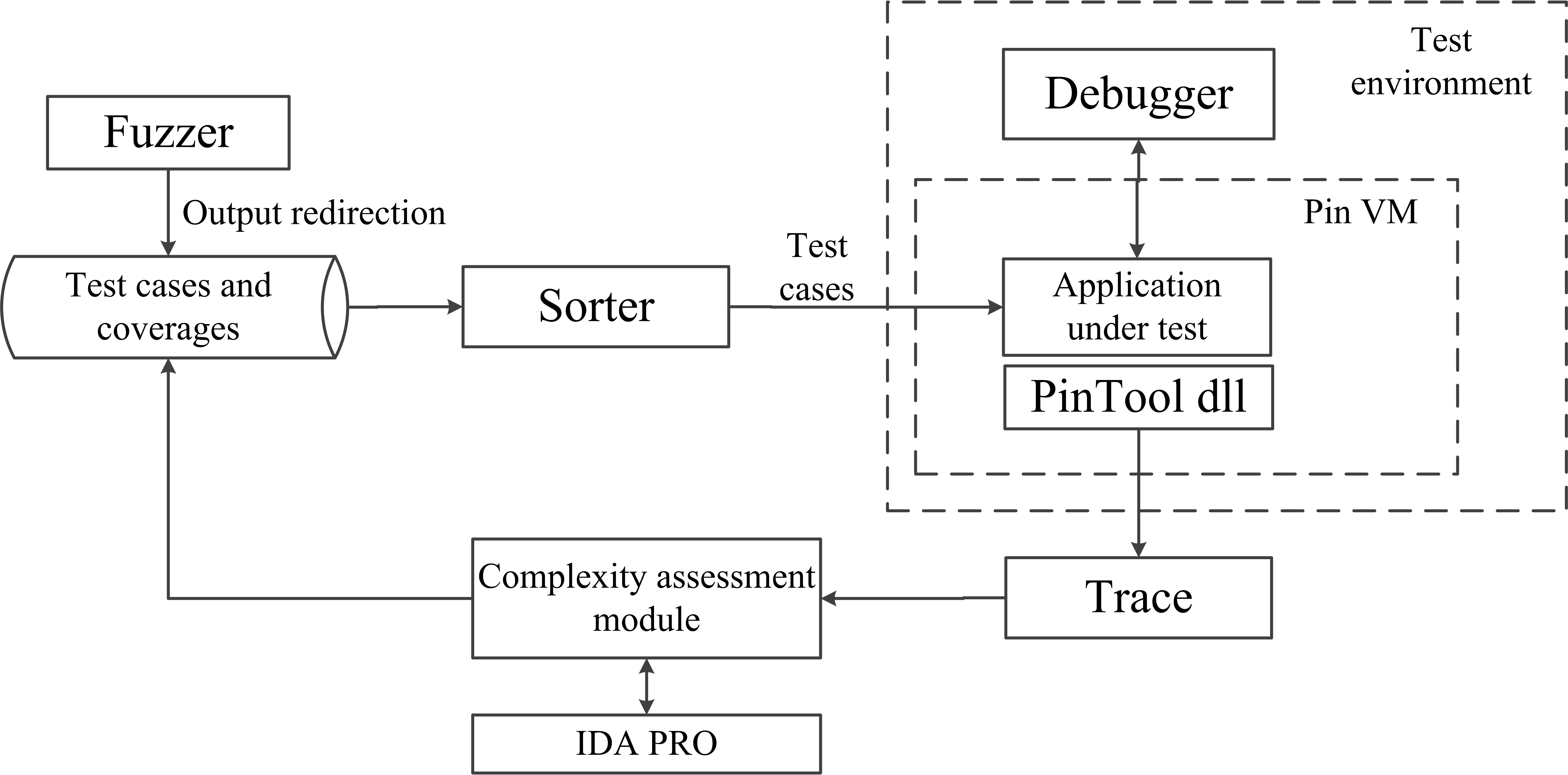}}
\caption{General scheme of the system}
\end{figure}

The output of the fuzzer is redirected first in the database to perform test cases prioritization according to fuzzing strategy. Then the system starts fuzzing and executable code instrumentation. For each test case the system evaluates new code coverage using obtained trace. Calculated coverages are written in the database (to use them further) and results are visualized on the screen. It should be noted that the process of complexity evaluation is performed in parallel with fuzzing to increase performance of the system. The tools were developed taking into account support of several platforms, thus making them easy to port across different operating systems with minimal changes.

\subsection{Experiments}

For experimental analysis of proposed approach, it was decided to estimate time costs for fuzzing campaign before and after integration of our system with 5 well-known fuzzers. We randomly selected 14 popular applications with known bugs from exploit-db, so as to include each type of bug that is considered in the article (stratification technique was used). Also we added 4 randomly selected applications (2 for Linux and 2 for Windows) from exploit-db with two and more bugs in one application to analyze capability of the system reduce time costs for several bug detections. Each software product was deployed in the private virtual environments within the following configurations: Windows 7 x64 (Intel Core i7 2.4 GHz with 2 Gb RAM), Windows Server 2008 SP2 x64 (Intel Core i7 2.4 GHz with 4 Gb RAM), Ubuntu Linux 12.10 (Intel Core i7 2.4 GHz with 4 Gb RAM). Experimental results presented in Figure 6.

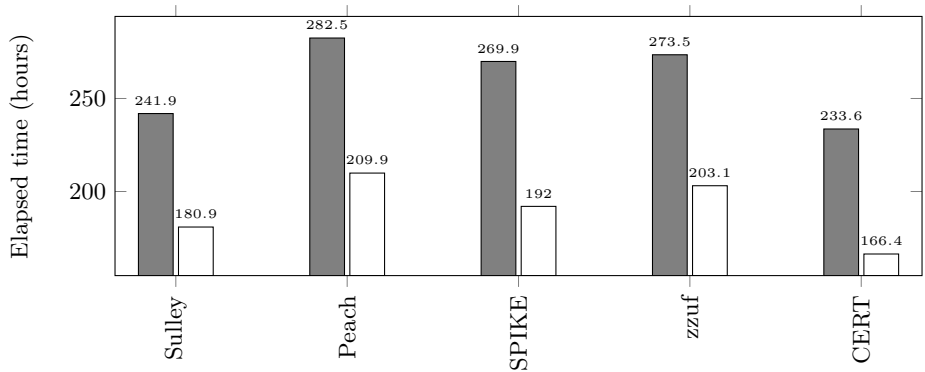
\begin{figure}
    \centering
	\begin{tikzpicture}
	\begin{axis}[
    ybar,
    ylabel = Elapsed time (hours),
    bar width= 13,
    enlarge x limits={abs=0.8cm},
	symbolic x coords={Sulley, Peach, SPIKE, zzuf, CERT},
	xtick=data,
    x tick label style={rotate=90},
    nodes near coords,
    every node near coord/.append style={font=\tiny},
    width=\linewidth,
    height = 0.26\textheight],
	\addplot[fill=gray] coordinates {(Sulley, 241.9) (Peach, 282.5) (SPIKE, 269.9) (zzuf, 273.5) (CERT, 233.6)};
	\addplot[] coordinates {(Sulley, 180.9) (Peach, 209.9) (SPIKE, 192.0) (zzuf, 203.1) (CERT, 166.4)};    
	\end{axis}
	\end{tikzpicture}
	\caption{The total time costs for fuzzing campaigns before and after integration of proposed system. The ordinate represents the total number of hours spent on testing all programs. White bar represents fuzzing campaign with proposed system. Zzuf \cite{L30}, CERT fuzzer \cite{L31}}
\end{figure}

\emph{Thus experimental results have shown that proposed system allowed to reduce time costs for testing by an average of 26-28\% for any considered fuzzer.} Detailed results may be found at \cite{L29}.

\section{Related works}

There are a lot of researches which performs fuzzing using some knowledge about testing application (white box fuzzing) to improve future tests generation, such as: symbolic execution or taint analysis \cite{L32}--\cite{L35}. Also, in several researches, authors try to use evolution algorithms \cite{L6}\cite{L36}\cite{L37} for effective data generation and increasing code coverage. Often, as an indicator of effectiveness is used the following metrics: the number of detected bugs, executed instructions, basic blocks and dangerous syscalls \cite{L6}--\cite{L9}\cite{L37}--\cite{L42}. Moreover, authors may apply special coverage criteria such as statements, decisions, and condition coverage \cite{L38}\cite{L39}\cite{L12}. In other case, researchers use input-based coverage criteria based on using input domain partitions and their boundary values \cite{L40}.

In some way, our approach has certain features in common with this paper \cite{L37}. Authors used a set of variables based on disassembly attribute information and application for procedure, such as the number and size of function arguments and local variables, the number of assembler code lines, procedure frame stack size and also cyclomatic complexity. In \cite{L41}, author uses cyclomatic complexity metric to perform in-memory fuzzing for more complex functions finding to increase a probability of bugs detection. In \cite{L12} author mentions about opportunity to apply cyclomatic complexity as a metric of effectiveness evaluation of the fuzzing technique. In \cite{L42} authors use basic blocks coverage to pick seed files to maximize the total number of bugs found during a fuzz campaign. In addition to coverage, they also consider other attributes, such as speed of execution, file size, etc. In \cite{L8} authors provide analysis of effective fuzzing strategies by using targeted taint driving fuzzing. Researchers used a different set of complexity metrics, such as cyclomatic complexity, attack surface exposure or static analysis for potentially vulnerable syscalls. The basic difference of our approach is that we use specially adapted metrics that take into account semantics of executed instructions as well as their interaction with input data.

\section{Discussion \& Future Work}

While implementing the metrics evaluation module, we limited ourselves to only general-purpose x86 instructions. Thus, in future, the module should also support co-processor group of instructions as well as applications for x64 and ARM architectures. Also we did not consider obfuscated executables since analysis of obfuscated code is a separate direction of research.

Secondly, we plan to start using metrics to automatically improve the efficiency level of data generation. For example, it makes sense to perform in-memory fuzzing for routines that have the highest level of complexity. It is also possible to generate data using evolutionary algorithms, in which we could use our set of efficiency assessment metrics as parameters for the data fitness function to improve data generation. Certainly, this approach needs to be confirmed experimentally.

It should be noted that the limitation of our approach is the fact that to reduce time costs, we need to have coverages array for each test case even before fuzzing. However if we do not have such coverages, reducing of time costs is only achieved at the second fuzzing campaign. This is justified when the system is being integrated within existed secure development life cycle \cite{L25}, when fuzzing is performed on the regular basis after new patch or functionality has been released. The system is also may be useful when existed set of test cases is applied for similar type of applications. Such fuzzing strategy makes sense, demonstrates positive results and is considered in this research \cite{L42}.

\section{Conclusion}

In this article, we propose the novel approach to reduce time costs of fuzzing campaign. We adapted 25 source code complexity assessment metrics to perform analysis in binary code. Our experiments on the 104 vulnerable applications have shown that Halstead B metric demonstrates maximum effectiveness to find vulnerable routines in comparison with other metrics. We also proposed our own metric based on Halsted B which shows more stable results. The experimental results of effectiveness assessment have shown viability of our approach and allowed to reduce time costs for fuzzing campaign by an average of 26-28\% for 5 well-known fuzzing systems. We have implemented our approach as a set of open-source tools that allows test cases prioritization, binary code complexity evaluation as well as performs code coverage analysis and results visualization.

This article is based upon work supported by the Russian Fund of Fundamental Research, research project №14-07-31350. This work was also  supported by the research grant for young Russian scientists 14.Z56.15.6012-MK.

\newpage
\section*{Appendix 1: Adapted Metrics List}

\newcommand*{\TitleParbox}[1]{\vspace{0.25cm} \parbox[c]{3cm}{\raggedright #1}}
\newcommand*{\FirstRowBox}[1]{\small\vspace{0.15cm} \parbox[c]{1cm}{\raggedright #1}}
\newcommand*{\LastParbox}[1]{\scriptsize \vspace{0.1cm} \parbox[c]{3cm}{\raggedright #1}\vspace{0.15cm}}

\begin{longtable}{c|c|c|c}
Metric & Symbol & Formula & Description \\
\multirow{4}{*}{Halstead metric} & H.V & $H.V = N \times log_2n$
                                        & \LastParbox{Program volume \newline
                                          $N = N_1 + N_2$ \newline
                                          $N_1$ - the total number of operators. \newline
                                          $N_2$ - the total number of operands. \newline
                                          $n = n_1 + n_2$ \newline
                                          $n_1$ - the total number of unique operators. \newline
                                          $n_2$ - the total number of unique operands.\newline
                                                      } \\
                                        & $H.N^*$
                                        & \small\TitleParbox{$H.N^* = n_1 \times log_2n_1 + n_2 \times log_2n_2$} 
                                        & \LastParbox{Calculated program length.} \\
                                        &  $H.D$
                                        & $H.D = \frac{n_1}{2} \times \frac{N_2}{n_2}$ 
                                        & \LastParbox{Program complexity.} \\
                                        &  $H.B.$
                                        & $H.B = \frac{E^\frac{2}{3}}{3000}$ 
                                        & \LastParbox{The number of delivered bugs.\newline$E=H.D \times H.B$} \\ \                                       
\FirstRowBox{Jilb's metric} & Jilb & $cl = \frac{CL}{n}$
                      & \LastParbox{
                        $CL$ - the total number of condition operators (jmp, jxx, etc.). \newline
                        $N$ - the total number of operators.
                        }
                        \\
\FirstRowBox{ABC metric} & ABC & $ABC = \sqrt{A^2 + B^2 + C^2}$ 
                  & \LastParbox{
                                 $A$ - assignments count. \newline
                                 $B$ - branches count. \newline
                                 $C$ - calls count. \newline
                                }
                                 \\
\FirstRowBox{Cyclomatic complexity} & CC & $CC = e - v + 2$ 
                                          & \LastParbox{$e$ – the number of edges;. \newline
                                                         $v$ - the number of nodes (basic blocks).
                                                        } \\ 
\FirstRowBox{Modified cycl. complex.} & CC\_mod & $CC\_mod = e - v^* + 2$ 
                           & \LastParbox{$v^*$ - the number of nodes (switch cases are considered as one node).
                                         } \\
\FirstRowBox{Pivovarskiy metric} & Pi & $Pi = CC\_mod + \sum_{i=0}^{n}p_i$ 
                          & \LastParbox{$p_i$ - nesting level of predicate node i. \newline
                                         $n$ - the total number of predicate nodes.}
                          \\
\FirstRowBox{Harrison \& Magel metric} & H\&M & $H\&M = \sum_{i=0}^{n}c_i$ 
                                           & \LastParbox{$c_i$ - node complexity. \newline
                                                          $n$ - the total number of predicate nodes.
                                                         } \\ 
\FirstRowBox{Boundary values metric} & Bound & $Bound = 1 - \frac{(n-1)}{S_a}$ 
                                         & \LastParbox{
                                                        $n$ - the total number of nodes. \newline
                                                        $S_a = \sum_{i}^{n}v_i$  - routine complexity \newline
                                                        $v_i = e_i-e_o$,\newline
                                                        $e_i$ - the total number of input edges. \newline
                                                        $e_o$ - the total number of output edges. \newline
                                                       } \\
\FirstRowBox{Span metric} & Span & $Span = \sum_{i=0}^{n}s_i$ 
                     & \LastParbox{$s_i$ - the number of statements containing the identifier. \newline
                                    $n$ - the total number of unique operators.
                                   } \\
\FirstRowBox{Henry \& Cafura metric} & H\&C & \TitleParbox{$H\&C= LOC \times (fan_{in} + fan_{out})^2$} 
                            & \LastParbox{$fan_{in}$ - the total number of input data flows. \newline
                                           $fan_{out}$ - the total number of output data flows.
                                          } \\
\FirstRowBox{Card \& Glass metric} & C\&G & $C\&G = S + D$ 
                            &\LastParbox{$S = fan_{out}^2$,
                                          $D =  \frac{v}{(fan_{out}+1)}$ \newline
                                          $v$ - the total number of input and output arguments.
                                         } \\
\FirstRowBox{Oviedo metric} & Oviedo & $Oviedo = \sum_{i=0}^{n}DEF(V_j)$ 
                        & \LastParbox{$DEF(V_j)$ - a number of occurrences of variable  $V_j$ from $R(i)$ set.\newline
                                       $n$ - a set of variables which is used in R(i).\newline
                                       $R(i)$ - a set of local variables defined in a node i first time.
                                      } \\
\FirstRowBox{Chapin metric} & Chapin & $Chapin = P + 2M + 3C$ 
                        & \LastParbox{P - the total number of output variables. \newline
                                       M - the total number of local variables. \newline
                                       C - the total number of variables which are used to manage CFG, such as: $cmp/test$ $var$ and then $jxx$.
                                      } \\
\FirstRowBox{Cocol metric} & Cocol & $Cocol = H.B + LOC + CC$ & \TitleParbox{} \\
\end{longtable}
\end{document}